\documentclass{Interspeech}

\interspeechcameraready

\title{Improving Practical Aspects of End-to-End Multi-Talker Speech Recognition for Online and Offline Scenarios}

\author[equalcontribution]{Aswin Shanmugam}{Subramanian}
\author[equalcontribution]{Amit}{Das}
\author[]{Naoyuki}{Kanda$^\dagger$}
\author[]{Jinyu}{Li}
\author[]{Xiaofei}{Wang}
\author[]{Yifan}{Gong}

\affiliation[nocounter]{}{Microsoft}{USA}
\email{aswins@microsoft.com}


\usepackage{comment}
\usepackage{cite}
\usepackage{amsmath,amssymb,amsfonts}
\usepackage{algorithmic}
\usepackage{graphicx}
\usepackage{textcomp}
\usepackage{amssymb}

\usepackage{xcolor}
\def\BibTeX{{\rm B\kern-.05em{\sc i\kern-.025em b}\kern-.08em
    T\kern-.1667em\lower.7ex\hbox{E}\kern-.125emX}}
\usepackage{multirow}

\renewcommand{\vec}[1]{\mathbf{#1}}

\newcommand{\uSOT}{uSOT }
\newcommand{\segSOT}{segSOT }

\newcommand{\CC}{ $<$cc$>$ }

\begin{document}

\maketitle

\ifinterspeechfinal
\def\thefootnote{$\dagger$}\footnotetext{This work was done while the author was at Microsoft, USA. The author is now with Meta, USA.}
\fi

\begin{abstract}    
We extend the frameworks of Serialized Output Training (SOT) to address practical needs of both streaming and offline automatic speech recognition (ASR) applications. Our approach focuses on balancing latency and accuracy, catering to real-time captioning and summarization requirements. We propose several key improvements: (1) Leveraging Continuous Speech Separation (CSS) single-channel front-end with end-to-end (E2E) systems for highly overlapping scenarios, challenging the conventional wisdom of E2E versus cascaded setups. The CSS framework improves the accuracy of the ASR system by separating overlapped speech from multiple speakers. (2) Implementing dual models—Conformer Transducer  for streaming and Sequence-to-Sequence for offline—or alternatively, a two-pass model based on cascaded encoders. (3) Exploring segment-based SOT (segSOT) which is better suited for offline scenarios while also enhancing readability of multi-talker transcriptions.
\end{abstract}

\vspace{-1mm}
\section{Introduction}
\vspace{-1mm}
End-to-end (E2E) automatic speech recognition (ASR) \cite{li2022recent, prabhavalkar2023end} has made significant strides in recent years, achieving remarkable performance on various benchmarks \cite{sainath2020streaming, Li2020Developing, radford2022robust}. However, the challenge of recognizing overlapping speech in multi-talker scenarios remains a critical area of research. Traditional ASR systems struggle with overlapping speech, leading to significant degradation in word error rates (WER) \cite{chen2020continuous,raj2020integration}. To address this, several approaches have been proposed, including permutation invariant training (PIT) \cite{pit,yu2017recognizing}, serialized output training (SOT) \cite{kanda_sot}, and continuous speech separation (CSS) \cite{yoshioka2018recognizing,chen2020continuous}.

SOT \cite{kanda_sot} was introduced to overcome some of the limitations of PIT. SOT uses a single output layer that generates transcriptions for multiple speakers sequentially, separated by a special token indicating speaker change. This approach eliminates the number of maximum speakers constraint and models dependencies among outputs for different speakers. However, SOT is primarily designed for offline ASR and does not support streaming inference. Token-level serialized output training (tSOT) \cite{kanda_tsot} further extends the SOT framework by generating recognition tokens for multiple speakers in chronological order, making it suitable for streaming ASR applications. 

CSS \cite{yoshioka2018recognizing,chen2020continuous} is another approach that has been explored for handling overlapping speech. CSS converts a long-form multi-talker speech signal into multiple overlap-free speech signals using a sliding window. Each of the separated signals can then be passed to a conventional single-speaker ASR system. While CSS has shown promise in improving ASR performance for overlapping speech \cite{yoshioka2018recognizing,chen2020continuous,chen2023speech}, it relies on a separate front-end processing step, which can introduce artifacts and errors that degrade the overall ASR accuracy.

Despite these advancements, current ASR systems still face challenges in balancing latency and accuracy, especially for practical applications that require both streaming and offline capabilities. Training separate models for each scenario is 
inefficient and introduces unnecessary complexity. Moreover, existing methods often struggle with highly overlapping speech and require complex architectures or multiple processing steps. 

To address these limitations, we propose three improvements in multi-talker ASR modeling. First, we leverage CSS single-channel front-end for E2E systems, challenging the conventional wisdom of E2E versus cascaded setups. The CSS single-channel front-end improves performance in highly overlapping scenarios by effectively separating speech from multiple speakers, thus enhancing the accuracy of the ASR system. 

While using an explicit front-end for multi-channel E2E multi-talker speech recognition has been shown to help \cite{chang2019mimo, subramanian2020directional, kanda_vararray}, such approaches naturally benefit from spatial filtering techniques like beamforming, which enable effective speech separation and improve recognition. In contrast, explicit speech separation for single-channel ASR has been less explored, as most E2E systems are trained directly for multi-talker recognition, relying on the model to implicitly learn both separation and transcription \cite{chang2020end}. We show that explicit speech separation using a single channel front-end provides significant advantages in highly overlapping scenarios compared to implicit separation within the ASR model.

Second, we implement dual models — Conformer Transducer (CT) \cite{gulati2020conformer, chen2021developing} for streaming and sequence-to-sequence (S2S) \cite{chorowski2014end, chan2016listen} for offline — and alternatively, a unified two-pass model based on cascaded encoders \cite{narayanan_cascaded} that balances accuracy with latency.
Finally, we explore segSOT ordering of multi-talker transcriptions to improve readability, turn-taking, and context for offline scenarios. We also study the effect of CSS encoder to further improve the accuracy of our offline model.

\section{Model}
\subsection{Conformer Encoder with CSS Input} \label{sec:tsot_css}
\begin{figure}[htbp]
\centerline{\includegraphics[scale=0.5]{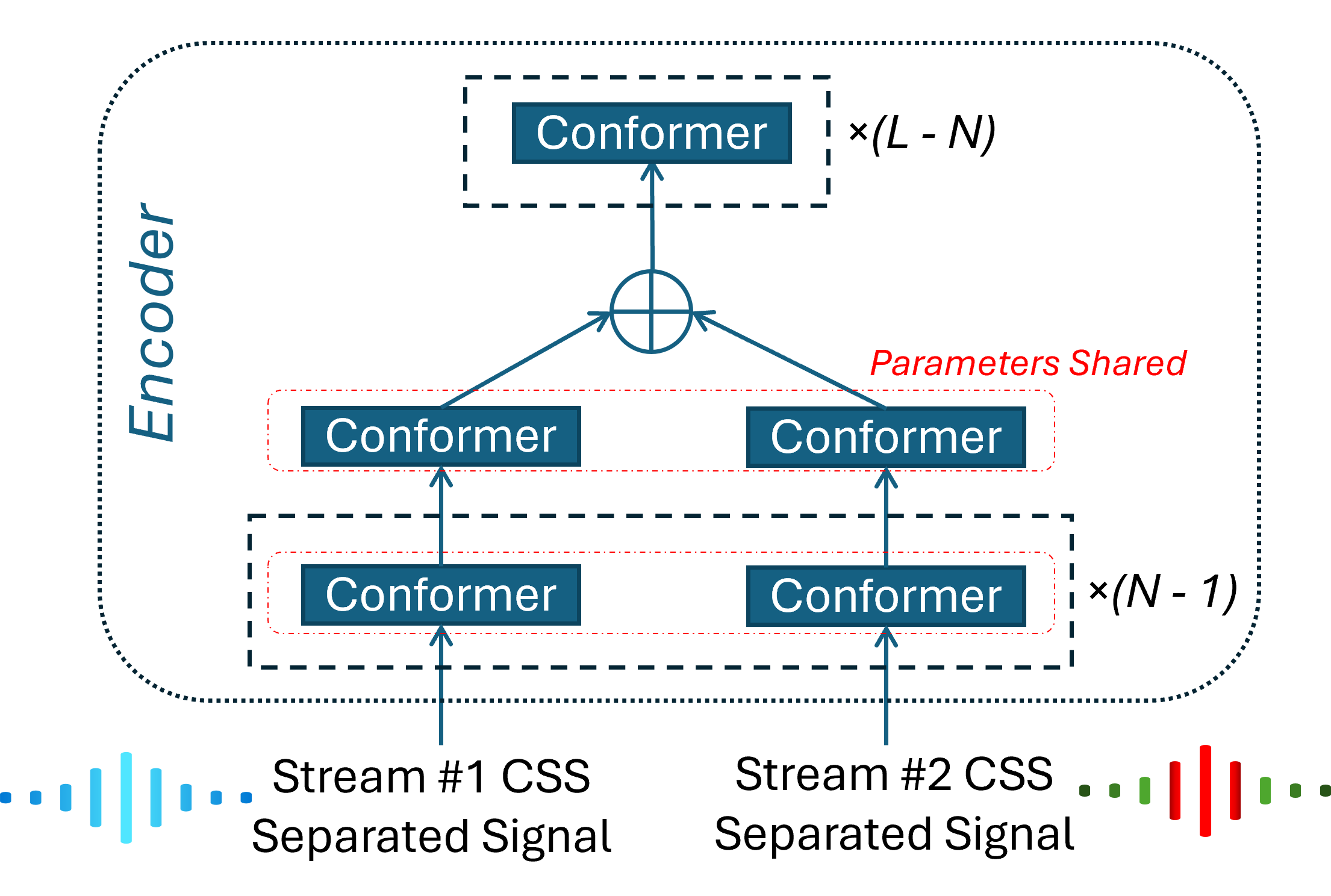}}
\caption{Two channel Conformer Encoder with CSS inputs}
\vspace{-4mm}
\label{fig:css_encoder}
\end{figure}

\vspace{-2mm}
We used an encoder architecture similar to what was used with a multichannel front-end in \cite{kanda_vararray}. As shown in Fig.~\ref{fig:css_encoder}, the encoder is designed to process two audio signals by splitting the conformer encoder with $L$ layers in total into $N$ channel dependent layers and $L-N$ channel independent layers.\footnote{Our study differs from \cite{kanda_vararray} in that we introduced single-channel CSS to enhance the single-channel multi-talker ASR model, whereas \cite{kanda_vararray} utilized multi-channel CSS.} The outputs from the N-th layer in each channel are summed and further processed by the $L-N$ channel independent layers. The parameters for the two-branched encoders are shared between the two channels. Consequently, the parameter count of the two-channel CSS Conformer Encoder is equivalent to that of the conventional single-channel non-CSS Conformer Encoder.

In the CSS front-end, a long-form audio input is segmented into overlapping chunks. Each chunk is then processed by a local Speech Separation (SS) network, which estimates two overlap-free signals (assuming that there are only two speakers in the chunk) that are input to the two-channel encoder in Fig.~\ref{fig:css_encoder}.  We use a conformer-based CSS \cite{chen2021continuous} similar to \cite{chen2023speech}. 

\subsection{Cascaded Encoder}
\vspace{-2mm}
\begin{figure}[tbp]
\centerline{\includegraphics[scale=0.5]{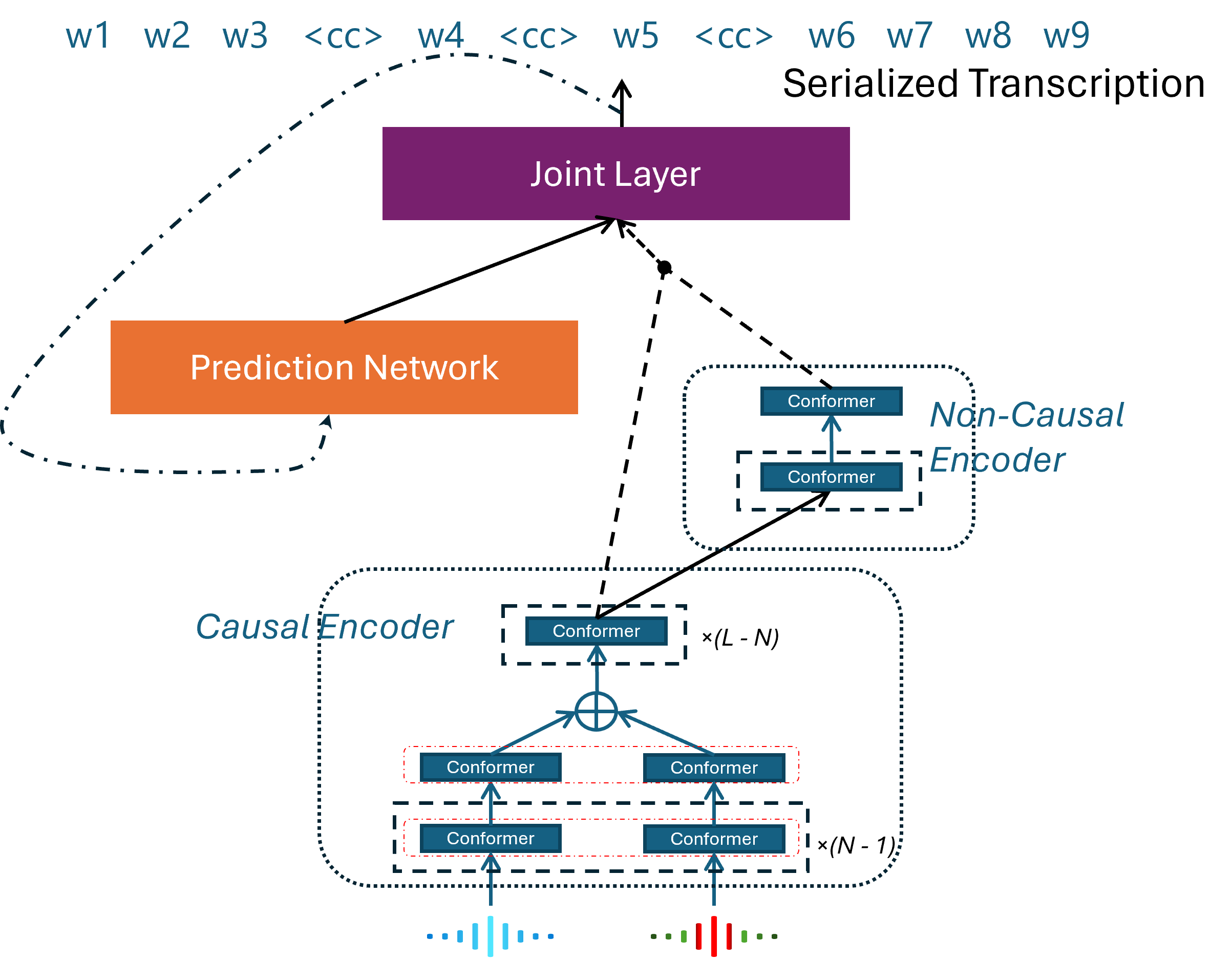}}
\caption{Conformer Transducer with Multi-Talker Cascaded Encoder. The Causal Encoder here takes two channel CSS inputs.}
\vspace{-5mm}
\label{fig:cascaded_ct_tsot}
\end{figure}
Cascaded encoders \cite{narayanan_cascaded} have previously been proposed to unify conventional streaming and non-streaming ASR models. The cascaded encoder model consists of a causal encoder and a non-causal encoder, where the causal encoder processes input features in a streaming fashion, and the non-causal encoder further processes these features using future context information. The outputs from both encoders are then fed into a shared Recurrent Neural Network Transducer decoder, allowing the model to operate in both streaming and non-streaming modes. 

Our proposed model shown in Fig.~\ref{fig:cascaded_ct_tsot}, leverages the strengths of the original cascaded encoder \cite{narayanan_cascaded} architecture while incorporating several key improvements to enhance performance in highly overlapping scenarios. We modify the causal encoder to consume two channel CSS inputs similar to Section \ref{sec:tsot_css}. We train it with multi-talker data and serialized output training. 

\subsection{S2S Model with Segment-based SOT (segSOT)}
\vspace{-2mm}
In this section, we explain the various components of the S2S-\segSOT model. It follows the architecture of an S2S model. Given a sequence of input acoustic frames  $\vec{x} = [\vec{x}_{1}, \vec{x}_{2}, \cdots, \vec{x}_{T}]$ and labels $\vec{y} = [y_{0}, y_{1}, \cdots, y_{U}]$, it models the distribution of the prediction labels conditioned on the entire input sequence $\vec{x}$ and the partial sequence of previously predicted labels, i.e., $P(y_{u}|y_{0}, y_{1}, \cdots, y_{u-1}, \vec{x})$ (non-causal auto-regressive). Since the predicted label at step $u$ is conditioned on the entire input sequence $\vec{x}$, it is suitable for modeling offline scenarios. 

\begin{figure}[t]

\includegraphics[height=28mm,width=\linewidth]{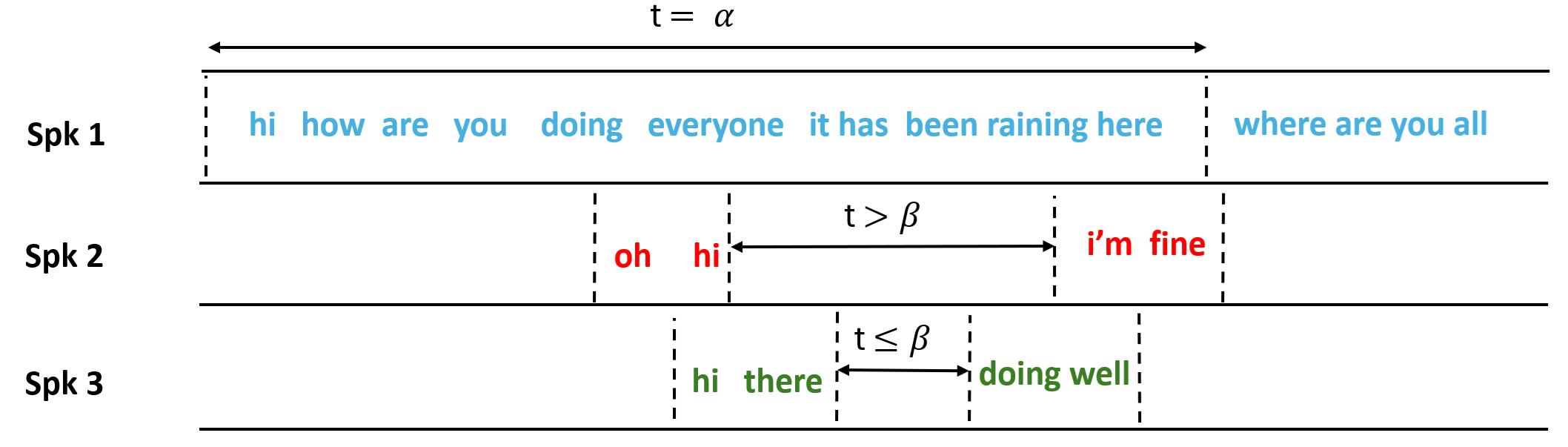}
\caption{Multi-talker Transcription ($\alpha$ = Maximum length of speech activity, $\beta$ = Maximum length of short pause).}
\vspace{-6mm}
\label{fig:multitalker_transcription}
\end{figure}

There are different ways of ordering/serializing the transcriptions in multi-talker simulations. One such way is the SOT paradigm \cite{kanda_sot, kanda2020joint}. Transcriptions of a multi-talker conversation are shown in Fig.~\ref{fig:multitalker_transcription}. There are 3 speakers with several regions of overlapped and non-overlapped speech. Ordering the transcriptions by start times of speakers and concatenating them yields an sSOT \cite{kanda_sot} transcription. Ordering by the start times of individual tokens/words yields tSOT \cite{kanda_tsot} transcriptions.  
 
We propose segSOT ordering of transcriptions which is suitable for offline scenarios. In segSOT, an utterance is split into \textit{segments} depending on speech activity or short pauses. The segments are then ordered according to their start times to yield a segSOT serialized transcription. 
The three different transcriptions (sSOT, tSOT, segSOT) corresponding to the scenario in Fig.~\ref{fig:multitalker_transcription} are shown below. \\ \\
\parbox[b]{1.0\linewidth}{\textbf{sSOT}: \textit{hi how are you doing everyone it has been raining here where are you all \CC oh hi i'm fine \CC hi there doing well}} \\ \\
\parbox[b]{1.0\linewidth}{\textbf{tSOT}: \textit{hi how are you  doing \CC oh \CC everyone \CC hi \CC hi \CC there \CC it has been \CC doing \CC raining \CC well \CC i'm \CC here \CC fine \CC where are you all}} \\ \\
\parbox[b]{1.0\linewidth}{\textbf{segSOT}: \textit{hi how are you doing everyone it has been raining here \CC oh hi \CC hi there doing well \CC i'm fine \CC where are you all}} \\

The \CC tag denotes a channel (or speaker) change. The length of a segment is determined by two parameters - a) $\alpha$: Maximum allowed length during speech activity and b) $\beta$: Maximum allowed length of a short pause. We design the maximum allowed length of a segment during speech activity to represent turn-taking scenarios. For example, in Fig.~\ref{fig:multitalker_transcription}, speaker 1 speaks continuously for more than $\alpha$ seconds. However, at t=$\alpha$, segSOT starts transcribing the earliest available segment of a different speaker, i.e., speaker 2 (\textit{oh hi}). This prevents large delays in transcribing other overlapping speakers thereby allowing frequent turn-taking scenarios which is common in multi-talker conversations. Following this, speaker 2 pauses for a while that exceeds $\beta$ seconds. Thus, segSOT stops transcribing speaker 2 and switches to the earliest available segment which is from speaker 3 (\textit{hi there}). Since there is a short pause ($\le \beta$), segSOT does not break the segment and continues to transcribe speaker 3 (\textit{doing well}). After this, it finds the earliest available segment which is from speaker 2 (\textit{i'm fine}). This process continues until all segments are exhausted.

There are several advantages of using segSOT.
\begin{itemize}

\item Readability: The readability of sSOT transcription can sometimes be difficult if there are several overlapping speakers. Although this can be improved with additional post-processing methods, the readability of segSOT transcriptions is much better since it is closer to the way humans transcribe, and no additional post-processing methods are required. 
\item Turn-taking and Context: There is utterance-based SOT (uSOT) \cite{kanda_usot} which orders the transcriptions according to the start times of the utterances. In \cite{kanda_usot}, full utterances were generated through prior segmentation of speech and text data. Later, those utterances were randomly sampled and overlapped to generate uSOT transcriptions. Since the length of the utterances were not properly defined, uSOT transcriptions are prone to large variations. For example, an utterance could encompass a long speech active region which impedes turn-taking among speakers. Alternatively, an utterance consisting of a long silence region flanked by two short speech active regions of unrelated text (eg. \textit{it is} $<$long sil$>$ \textit{got it}) results in an incoherent transcription (\textit{it is got it}). However, with segSOT, these problems are avoided thereby achieving better consistency in  turn-taking and context through choices of $\alpha$ and $\beta$ parameters. Furthermore, the sSOT ordering precludes turn-taking during overlapped speech and \cite{kanda_usot} has shown that sSOT ordering performed worse than \uSOT ordering.
\item CTC: Since S2S models are vulnerable to hallucinations, an auxiliary CTC objective criterion is usually added to mitigate this problem \cite{kim2017joint}. While word ordering in sSOT/uSOT is more non-monotonic than segSOT, the CTC criterion favors monotonicity. Because of this conflict, the CTC objective tends to penalize sSOT/uSOT more severely than segSOT. 
\end{itemize}

\section{Experiments}
\vspace{-2mm}
\subsection{Data}
\vspace{-2mm}
We trained two seed single-speaker ASR models (CT, S2S) using 30,000 hours of in-house data \cite{li2020high}, with all personally identifiable information removed. To develop multi-speaker models, we fine-tuned the initial seed model using a diverse dataset. This dataset included: a) simulated multi-speaker data derived from the aforementioned 30,000 hours of recordings, b) real meetings data from the AMI \cite{carletta2006amimeeting} and ICSI \cite{janin2003icsimeeting} corpora, and c) some in-house meeting recordings. During the multi-speaker simulation, two utterances were randomly mixed in approximately two-thirds of the cases, while the remaining one-third consisted of the original single-speaker utterances.

We evaluated our models using the LibriCSS test set \cite{chen2020continuous}. The original recordings were made with a 7-channel microphone array, but we used only the first channel for our experiments. The recordings span a total of 10 hours and are categorized by speaker overlap ratios ranging from 0\% to 40\%. Each category includes 10 mini-sessions, each lasting 10 minutes. For our evaluation, we used the segmented versions of sessions 1 through 9, excluding session 0. We measure the performance with Speaker Agnostic Word Error Rate (SAgWER) computed using NIST's \texttt{asclite} \cite{nistsctk} tool.

\begin{table*}[h]
  \centering
  \caption{SAgWER (\%) on the monaural LibriCSS test set. A macro average of SAgWERs is shown in
the “Avg.” column. 0L and 0S are 0\% overlap conditions with long and short inter-utterance silences. Column ``CSS" represents whether the two-channel CSS Encoder is enabled or not. Column ``C,NC" represents the number of causal and non-causal encoder layers. T is a variable representing the length of the input utterance group.}
    \vspace{-3mm}
    \begin{tabular}{c|c|c|c|c|c|ccccccc}
    \hline
    \multirow{2}[0]{*}{ID} & \multirow{2}[0]{*}{System} & \multirow{2}[0]{*}{CSS} & \multirow{2}[0]{*}{C,NC} & \multirow{2}[0]{*}{Latency} & \multirow{2}[0]{*}{Mode} & \multicolumn{7}{c}{SAgWER (\%) for different overlap ratios} \\ 
      &    &  &  &  & & 0L    & 0S    & 10    & 20    & 30    & 40    & Avg. \\ \hline
    1 & CT-tSOT  & $\times$ & 18,0 & 160ms & Online & 7.76  & 7.33  & 8.73  & 11.78 & 14.25 & 16.47 & 11.52 \\
    2 & CT-tSOT &  \checkmark & 18,0 & 160ms & Online & 8.18  & 8.76  & 9.55  & 11.62 & 13.07 & 14.97 & 11.38 \\ \hline
    3 & Cascaded CT-tSOT & \checkmark & 12,6 & 160ms & Online & 11.34 & 11.79 & 12.51 & 15.22 & 16.83 & 19.11 & 14.87 \\    
    4 & Cascaded CT-tSOT & \checkmark & 12,6 &  5s & Offline   & 7.23  & 7.61  & 8.54  & 10.09 & 11.80 & 13.97 & 10.22 \\
    5 & Cascaded CT-segSOT & \checkmark & 12,6 & 5s & Offline  & 9.84  & 8.03  & 8.60  & 11.07 & 13.29 & 16.58 & 11.56 \\ \hline 
    6 & S2S-segSOT  & $\times$ & 0,18 & T & Offline & 7.05  & 7.23  & 8.15  & 9.93 & 11.96 & 14.84 & 10.24 \\
    7 & S2S-segSOT &  \checkmark & 0,18  & T & Offline & 7.30  & 7.46  & 8.01  & 9.53 & 11.30 & 14.01 & 9.93 \\ \hline
    \end{tabular}%
  \label{tab:libricss}%  
  \vspace{-4mm}
\end{table*}%

\vspace{-2mm}
\subsection{Model}
\vspace{-1mm}
\subsubsection{CSS Front-end}
\vspace{-1mm}
The training of the CSS network utilizing WavLM~\cite{chen2022wavlm} follows the methodology outlined in \cite{chen2023speech, wang2022handling}, which adopts a sliding window approach (2.4s window + 0.8s hop). Initially, WavLM is trained using unlabeled speech data. Subsequently, the conformer-based SS network \cite{chen2021continuous} is fine-tuned. This process involves inputting both an audio spectrogram and the WavLM embedding into the network. The WavLM embedding itself is derived from the weighted average of the outputs across all transformer layers of WavLM, with weights adjusted during fine-tuning. The optimization of the SS network employs an utterance-level permutation invariant training loss for two outputs, where each output branch is tasked with estimating a magnitude mask for each speaker.  This design introduces a processing latency of 0.8s. In \cite{wang2022handling}, the RTF is calculated as 0.548 for this frontend. The overall latency of our proposed system is the maximum of this 0.8s and the latency of ASR. 

We trained our ASR models without and with the two-channel CSS encoder. ASR models without the CSS encoder were trained with 80-dimensional log Mel filter bank (LMFB) features extracted every 10ms from a window of 25ms of speech to handle the  mixed band 8kHz or 16kHz speech with the method described in \cite{li2012improving}. ASR models with the two-channel CSS encoder were trained with 80-dimensional LMFB features extracted from each channel. 
\vspace{-2mm}
\subsubsection{CT-tSOT Model and Cascaded CT Model}
\vspace{-1mm}
The CT model encoder comprises 2 convolutional layers followed by an 18-layer Conformer encoder, utilizing a chunk-wise streaming mask to achieve a latency of 160ms. Each Conformer block includes a multi-head self-attention (MHSA) layer with an attention dimension of 512, distributed across 8 heads, and a 2048-dimensional feed-forward network (FFN) with a Gaussian error linear unit (GELU). 

For the cascaded CT model we used 12 causal layers and 6 non-causal layers so that overall parameters are the same as the CT model. The non-causal layers have a chunk size of 5s. The prediction network for the transducer models consists of a 2-layer LSTM with 1024 dimensions, and the joint network is a 512-dimensional feed-forward layer. We utilized 4,003 tokens for recognition, along with blank and $\langle \text{cc} \rangle$ tokens. We used the AdamW optimizer with $(\beta_1, \beta_2) = (0.9, 0.98)$ and a peak learning rate (PLR) of $2.0 \times 10^{-4}$. The LR schedule followed a linear decay in LR from the PLR over 4 million steps.

\vspace{-2mm}
\subsubsection{S2S-\segSOT Model}
\vspace{-1mm}
The S2S-\segSOT model consists of an 18-layer Conformer encoder and a 6-layer transformer decoder. The architecture of Conformer encoder blocks are the same as the CT model. Each decoder block consists of an MHSA layer with an attention dimension of 512, distributed across 8 heads, and a 2048-dimensional FFN followed by Rectified Linear Unit (RELU). 

The S2S-\segSOT model was trained with a combination of label smoothing and CTC auxiliary losses using weights of 1 and 0.2 respectively on segSOT transcriptions. The segSOT transcriptions were generated using $(\alpha, \beta) = (5, 0.5)$ seconds. The same AdamW optimizer of the CT model was used but with a PLR of $2.2 \times 10^{-5}$. The LR schedule followed a linear increase in LR up to the PLR over 300k warmup steps followed by a linear decay up to 3 million steps. 

\vspace{-2mm}
\subsection{Results}
\vspace{-2mm}
The evaluation results for our models tested on LibriCSS are tabulated in Table~\ref{tab:libricss}. It should be noted that no Librispeech data was used during the training. Hence, the results represent the performance of our models in unseen test conditions. CT-tSOT is the baseline in our study and it was shown in \cite{kanda_tsot} to perform similarly or better than using a cascaded system (CSS + single-talker ASR), even with only simulated training data. Here, we also use real overlapped multi-talker data in training, which benefits E2E multi-talker systems. Training single-talker ASR with such real data is not straightforward as we cannot obtain channel-specific error-free target transcriptions even if we preprocess it with CSS.

First, in rows 1-2, we compare the performance of the multi-talker streaming CT-tSOT system with and without the proposed CSS encoder. The zero overlap test sets in LibriCSS (``0L", ``0S") represent single-talker test sets. For both the test sets, there was some degradation when using the CSS front-end. However, there was significant improvement (8-9\% relative) in the highly overlapping scenarios (``30", ``40") with the overall average across all scenarios showing a marginal improvement. 

In rows 3-5, we discuss the results of the cascaded CT model with the CSS encoder. Row 3 shows the first pass (streaming) results of the cascaded CT-tSOT model where only the first 12 layers of causal encoders were used. Non-causal encoders were not used at all. The results are significantly worse than row 2 as there are fewer number of causal encoder layers, i.e., 18 in row 2 versus 12 in row 3. Our cascaded system was designed with fewer number of causal encoder layers to match the total number of parameters with the CT systems in rows 1-2. Next, in row 4, we compare the second pass results of the cascaded CT-tSOT model using both the causal and non-causal encoder layers. Clearly, this model significantly outperforms the CT-tSOT model in row 2 across all scenarios. The improvement is 10.2\% relative (11.38 $\rightarrow$ 10.2) at the cost of more latency. This demonstrates a trade-off between accuracy and latency within the same model using the first pass and second pass outputs. Using the segSOT transcription ordering with cascaded CT model (row 5) causes some degradation as the segSOT ordering makes the alignment more complicated with RNN-T loss but improves on readability.

Next, we discuss the results of the proposed purely offline S2S models. %First, we compare the proposed S2S-\segSOT model with S2S-\uSOT model. The average WERs of S2S-\uSOT and S2S-\segSOT models (both without CSS encoder) are 11.36 and 10.24 respectively establishing the superiority of S2S-\segSOT modeling.
To assess the performance of our models in single-talker test cases, we compare the multi-talker S2S-\segSOT model with a baseline single-talker S2S model. Both these models were trained the same number of parameters. The results are: S2S $\rightarrow$ segSOT: 7.02 $\rightarrow$ 7.05 (0L test) and 7.15 $\rightarrow$ 7.23 (0S test). There is some degradation but it is quite marginal ($\le$ 1\%). This shows that the multi-talker model is able to preserve the capacity of the single-talker model.

Finally, in rows 6-7, we compare the S2S-\segSOT model with CSS encoder (row 7) and without CSS encoder (row 6). The proposed S2S-\segSOT model without the CSS encoder performed better than the model with CSS encoder in single-talker scenarios (0L, 0S) by only about 3.3\% relative. Moreover, the model tends to perform better with long utterances (0L: 7.05, 7.30) than with short utterances (0S: 7.23, 7.46) because of longer contexts. However, for the more challenging scenarios of overlapped speech (10-40), the model equipped with CSS encoder (row 7) performed better on an average by about 4.2\%  relative. The robustness of S2S-\segSOT model with CSS encoder is further highlighted by the fact that larger the amount of overlapped speech, the wider the performance gap with respect to the non-CSS encoder model. Moreover, it achieves the highest accuracy (9.93\%) among all the models in Table~\ref{tab:libricss}.

Finally, we compare the best Cascaded CT-tSOT model (row 4) and with S2S-\segSOT model (row 6). We observe that the Cascaded CT-tSOT model with CSS encoder, despite limited latency, is able to achieve on par accuracy (10.24\%) with the S2S-\segSOT model which is a purely offline model.

\vspace{-2mm}
\section{Conclusion}
\vspace{-2mm}
In this paper, we have proposed three improvements in E2E multi-talker ASR modeling that effectively balances latency and accuracy for both streaming and offline scenarios. First, we leverage speech separated signals by using two channel CSS encoder in our ASR models. Second, we implemented online and offline models and introduced a unified two-pass model based on cascaded encoders. Finally, we explore segSOT to improve modeling multi-talker conversations for offline scenarios. We demonstrated that a combination of CSS encoder and cascaded model, despite limited latency, is able to achieve high degree of accuracy similar to that of S2S-segSOT which is a purely offline model. We also showed that CSS encoder tends to improve the offline S2S-segSOT  model yielding the best performance. Overall, our work advances the state-of-the-art in multi-talker speech recognition, providing a practical solution for real-time captioning and summarization applications. 
\newpage
\bibliographystyle{IEEEtran}
\bibliography{refs}

\end{document}